\documentclass[pra,preprint,superscriptaddress,showpacs]{revtex4}

\usepackage{graphics}           
\usepackage{bm}                 
\usepackage{amsmath}            
\usepackage{amssymb}
\usepackage{verbatim}           
\usepackage{amsthm}             
\theoremstyle{plain}            

\def\ket#1{{|#1\rangle}}
\def\bracket#1#2{{\langle#1|#2\rangle}}

\begin{document}

\title{Bounds on QCA Lattice Spacing from Data on Lorentz Violation}

\author{Leonard \surname{Mlodinow}}\email{lmlodinow@gmail.com}
\author{Todd A. \surname{Brun}}\email{tbrun@usc.edu}
\affiliation{Center for Quantum Information Science and Technology, University of Southern California, Los Angeles, California}

\date{\today}

\begin{abstract}
Recent work has demonstrated that discrete quantum walks, when extended to quantum cellular automata (QCA), can, in the continuum limit, reproduce relativistic wave equations and quantum field theories (QFTs), including free quantum electrodynamics (QED). This QCA/QFT correspondence bridges quantum information processing and high-energy physics, raising fundamental questions about the nature of spacetime: whether it is the continuum QFT or the discrete QCA that is fundamental. For while Lorentz invariance appears robust experimentally, it may only approximate a deeper discrete structure, particularly at Planck-scale energies. This high-energy Lorentz violation is potentially observable either through cumulative effects over cosmic distances or via small deviations at accessible energies. In this paper, we analyze the QCA corresponding to QED and show that it implies both a deviation from the speed of light and spatial anisotropies. Using current experimental and astrophysical constraints, we place upper bounds on the QCA lattice spacing, providing insight into the plausibility of a fundamentally discrete spacetime.
\end{abstract}


\maketitle

\section{Introduction}

Discrete quantum walks---built on simple principles and symmetries---can, in the continuum limit, yield relativistic wave equations such as the Weyl, Dirac, or Maxwell equations. This correspondence has been extended to quantum field theories (QFTs) by constructing a quantum cellular automaton (QCA) from the quantum walk \cite{Bisio15a,Bisio15b,Mallick16,MlodinowBrun20,BrunMlodinow20,MlodinowBrun21,Arrighi20,Eon23,Centofanti24,Bisio25}. Most notably, it has been demonstrated that free quantum electrodynamics (QED) can emerge from the continuous limit of lattice-based Fermi and Bose QCA models \cite{MlodinowBrun21,BrunMlodinow25}.

This QCA/QFT correspondence is striking in that it connects two traditionally distinct branches of physics: quantum information processing, which involves the action of a sequence of quantum gates (unitary operations) on an initial state of a set of quantum systems (e.g., qubits); and quantum field theory, which involves the time development of a quantum field given by the action of a unitary operator on a state describing quantum particles or the creation and annihilation operators corresponding to them. But apart from such theoretical considerations, the correspondence raises a fundamental question: Is the QCA merely a model that in the continuum limit reproduces the QFTs that describe nature, or is the QFT merely a small-lattice-spacing approximation to the QCA theory, which is what corresponds to nature? 

Although we typically assume that QFT is the ``real'' theory, it is experimentally impossible to distinguish between continuous and discrete models if the underlying lattice spacing is small enough. If spacetime is fundamentally discrete, Lorentz symmetry would emerge as approximation, valid at long length scales, and the role of experimental and observational tests of Lorentz invariance would be to place bounds on the maximum lattice spacing. 

The idea that Lorentz symmetry might be violated at very high energies and very short length scales is nothing new---Dirac, for example, proposed in the 1950s that Lorentz invariance violation might play a role in physics \cite{Dirac51}, followed by several others in the ensuing years \cite{Bjorken63,Pavlopoulos67,Redei67}. In the 1990s, influential papers by Coleman and Glashow raised the subject of systematic tests of Lorentz violation within the context of elementary particle physics \cite{Coleman97,Coleman99}. Today, many theories of quantum gravity also predict Lorentz violation---for example, string theories \cite{Kostalecky89a,Kostalecky91}, loop quantum gravity \cite{Gambini99,AmelinoCamelia13,Mavromatos09,KowalskiGlikman05,Myers03}, theories of non-commutative geometry \cite{Hayakawa00,Carroll01}, and brane-world scenarios \cite{Burgess02a,Frey03,Cline04}. In fact, it has been conjectured that a violation of Lorentz Invariance is an important observable signature of Quantum Gravity \cite{Gambini99,Kostelecky89b,Ellis99,Burgess02b}.

In quantum gravity theories the natural scale at which one would expect to observe Lorentz violation is the Planck energy of approximately $10^{19}$ GeV. This corresponds to a distance scale of the Planck length $L_P = 1.6\times 10^{-35}$ m, so one might consider this to be a natural scale for the QCA lattice. Unfortunately, the Planck energy $E_P$ is not just far higher than current accelerator energies of approximately $1.4\times10^4$ GeV, but also far higher than the energy of the most energetic observed particles: ultra high-energy cosmic rays that have energies as high as $10^{11}$ GeV are still less energetic than $E_P$ by a factor of $10^8$. However, large violations of Lorentz invariance at the Planck scale can lead to a small degree of violation at much lower energies, presenting the possibility of detection at more realistic scales. Alternatively, the deviations can accumulate to detectable levels over large propagation distances making them amenable to astrophysical observation. As a result, the past few decades have seen a growing literature on tests of Lorentz invariance, and on the placement of bounds on a variety of proposed deviations \cite{Mattingly05,Liberati13,Flambaum17,Ellis19,Cao24,Desai24,Piran24}. Several recent papers have even reported finding some evidence of it \cite{Zhang15,Finke23,Galanti25}.

In the current paper we examine the discrete lattice QCA that corresponds to QED in the continuum limit. We show that it predicts both a deviation of the speed of light from $c$, and a degree of anisotropy at high energies. We use current experimental and observational data that constrains the magnitude of such deviations to place an upper limit on the lattice size in the QCA theory.

\section{Observational limits on Lorentz Invariance Violation and anisotropy}
\label{sec:observation}

\subsection{Lorentz Invariance Violation}
\label{sec:LIVexpts}

Though the Planck scale (roughly $10^{19}$ Gev) is often considered the natural energy threshold for Lorentz Invariance Violation (LIV), it has long been conjectured that signatures of LIV might be observable at much lower energies. One such signature is an energy-dependent modification to the vacuum dispersion relation for photons \cite{AmelinoCamelia98}:
\begin{equation}
E^2 \approx p^2 c^2 \left[ 1 - \sum_{n=1}^\infty s \left( \frac{E}{E_{\mathrm{QG},n}} \right)^n \right] ,
\label{eq:dispersion}
\end{equation}
Here $E$ and $p$ represent the photon's energy and momentum, respectively; $s = \pm1$ indicates the sign of the LIV correction (corresponding to either subluminal or superluminal propagation); and $E_{\mathrm{QG},n}$ is called the quantum gravity energy scale, a parameter to be determined, or bounded, through observation and experiment.

At energies much smaller than the quantum gravity scale $E\ll E_{\mathrm{QG},n}$, the infinite series in the modified dispersion relation is effectively governed by its lowest-order terms. As a result, only the first two terms---specifically, those with $n=1$ (linear) and $n=2$ (quadratic)---are relevant for phenomenological studies of LIV.

To get an expression for the photon group velocity from Eq.~(\ref{eq:dispersion}) we take the square root, expand in powers of $E/E_{\mathrm{QG},n}$, and then substitute the leading order $pc$ for $E$ in the correction term on the right-hand side to get:
\begin{equation}
E \approx pc \left[ 1 - \frac{s}{2} \left( \frac{pc}{E_{\mathrm{QG},n}} \right)^n \right] ,
\end{equation}
(where $n=1$ or $n=2$, depending on the model). The group velocity is thus:
\begin{equation}
v(E) = \frac{\partial E}{\partial p} \approx c\left[ 1 - \frac{s}{2} (n+1) \left(\frac{pc}{E_{\mathrm{QG},n}} \right)^n \right] \approx c\left[ 1 - s_\pm \frac{n+1}{2} \left(\frac{E}{E_{\mathrm{QG},n}} \right)^n \right] .
\label{eq:groupvelocity1}
\end{equation}

This dependence of the speed of photons on their energies means that two photons emitted simultaneously from a distant astronomical source with different energies would not reach an observer at the same time. Although the effect is extremely small, over cosmological distances it could build up to a measurable time delay. Accordingly, extensive efforts have been made to detect LIV signatures by examining spectral lags---the time differences in the arrival of high- and low-energy photons, where a positive lag indicates that the higher-energy photons arrive first \cite{Bolmont20}.

The strongest constraints to date on the quantum gravity energy scale $E_{\mathrm{QG},n}$ are derived from time-of-flight measurements of more than 64,000 photons in the energy range of 0.2--7 TeV from GRB 221009A, emitted within the first 4000 s after the MeV burst trigger, and detected by the Large High-Altitude Air Shower Observatory (LHAASO) \cite{Cao24,LHAASO23}. These high-precision measurements of GeV photons have placed robust bounds on potential LIV-induced vacuum dispersion. In the case of linear corrections, the limits are $E_{\mathrm{QG},1} > 1 \times 10^{20}$ GeV for the subluminal scenario and $E_{\mathrm{QG},1} > 1.1 \times 10^{20}$ GeV for the superluminal case \cite{Cao24}. For quadratic modifications, the corresponding constraints are $E_{\mathrm{QG},2} > 6.9 \times 10^{11}$ GeV (subluminal) and $E_{\mathrm{QG},2} > 7.0 \times 10^{11}$ GeV (superluminal).The linear constraints are particularly significant because they reach, and even slightly surpass, the Planck energy scale.

The absence of detectable energy-dependent time delays within the sensitivity of LHAASO excludes a wide range of LIV models predicting substantial linear or quadratic effects at or below these energies. Consequently, GRB 221009A stands as a reference point in the ongoing effort to detect Lorentz violation in high-energy astrophysical phenomena. Continued and future observations of gamma-ray bursts, active galactic nuclei, and pulsars are expected to further refine these bounds---or possibly reveal anomalies that could signal new physics beyond the standard model and general relativity.

As we will show in Sec.~\ref{sec:QW}, for our QCA massless boson model \cite{MlodinowBrun21,BrunMlodinow25a}, $s$ can be either positive or negative, and the leading correction to the speed of light corresponds to $n=1$, yielding
\begin{equation}
v(p) \approx c - s\Delta v(p) ,
\end{equation}
where $\Delta v(p)$ is proportional to the lattice spacing $\Delta x$, enabling us to translate the observational bounds on $\Delta v(p)$ to bounds on $\Delta x$.

\subsection{Anisotropy}
\label{sec:anisotropyexpts}

Discrete models of QED typically predict not just a deviation from the magnitude of the speed of light at high energies, but also a violation of isotropy: while no axis may be preferred to another, in these models light travels at a different speed along an axis than at an angle relative to them. This is also true of our massless boson QCA model. As we will see, the correction to the speed of light is direction-dependent; wave packets propagate along the lattice axes with constant velocity $c$, but otherwise with an energy-dependent correction that can be positive or negative, depending on the direction:
\begin{equation}
v(p) \approx c - s(\theta,\phi) \Delta v(p) ,
\label{eq:groupvelocity2}
\end{equation}
where $\theta$ and $\phi$ are angular coordinates:
\begin{equation}
v_x = v\sin(\theta)\cos(\phi) , \ \ \  
v_y = v\sin(\theta)\sin(\phi) , \ \ \  
v_z = v\cos(\theta) .
\label{eq:vspherical}
\end{equation}

The most precise experimental constraints on such anisotropy have been obtained through modern incarnations of the Michelson-Morley experiment \cite{MichelsonMorley1887}. These experiments compare resonance frequencies in optical \cite{Mueller03,Hermann05,Antonini05,Hermann06,Herrmann09,Eisele09,Nagel15} or microwave \cite{Wolf04,Stanwix06} cavities, which are either rotated mechanically on turntables or analyzed as the Earth rotates. The most stringent bounds were reported by \cite{Herrmann09,Eisele09,Nagel15}, setting a limit on an anisotropy of the speed of light of $\Delta c/c \lesssim 1\times 10^{-17}$--$10^{-18}$. (A more stringent bound is reported in \cite{Flambaum17}, but this relies on an assumption that the speed of matter particles---protons and neutrons---is bounded by $c$, which is not necessarily true for QCA models).

We will compare these experimental limits on anisotropy to the predictions of our QCA model in Sec.~\ref{sec:bounds}, which will give an additional upper bound on the size of the lattice spacing. These limits are less tight than those arising from tests of LIV from gamma ray bursts (GRBs), because they use much lower-energy photons; however, they do not suffer from the issue that gamma ray bursts are relatively rare events, and their directions are not under experimental control.

\section{The QW for massless bosons}
\label{sec:QW}

Since in this paper we are considering a single particle, we do not need the full machinery of a QCA. Instead, we consider a quantum walk on a lattice, which corresponds to the single-particle sector of a QCA. We consider the cubic lattice, comprising a set of vertices $\mathbf{x} \equiv (x,y,z) = (i\Delta x, j\Delta x, k\Delta x)$, where $i, j, k$ are integers and $\Delta x$ is a fixed lattice spacing. A discrete-time walk on this lattice involves moving $\Delta x$ along the three cardinal directions at each time step. The evolution from one time to the next is given by a unitary evolution operator $U_{\rm QW}$ of the form \cite{Chandrashekar11,Chandrashekar13,MlodinowBrun18}
\begin{equation}
U_{\rm QW} = e^{i\theta Q} \left(S_X P^+_X + P^0_X + S^\dagger_X P^-_X \right) \left(S_Y P^+_Y + P^0_Y + S^\dagger_Y P^-_Y \right) \left(S_Z P^+_Z + P^0_Z + S^\dagger_Z P^-_Z \right) .
\label{eq:bosonQW}
\end{equation}
Here, the projectors $P^{+,-,0}_{X,Y,Z}$ are projectors onto the internal space of the particle, and $S_{X,Y,Z}$ are the shift operators that move the particle by $+\Delta x$ along the corresponding axis, and $S^\dagger_{X,Y,Z}$ moves by $-\Delta x$. The zero mode projector $P^0_{X,Y,Z}$ corresponds to the particle staying in place along that axis. The operator $Q$ is the ``coin flip'' operator, that also acts on the internal space. In earlier work \cite{MlodinowBrun21,MlodinowBrun18} we found that the ``coin flip'' parameter $\theta$ ends up playing the role of a mass in the long-wavelength limit. For a massless boson, therefore, we can choose $\theta=0$ and we don't have a coin-flip operator $Q$.

For a single axis, say $X$, the three projectors $P^{+,-,0}_X$ are orthogonal and add up to the identity:
\[
P^i_X P^j_X = \delta_{ij} P^i_X , \ \ i,j=+,-,0,\ \ \ \sum_i P^i_X = I .
\]
But projectors corresponding to different axes are not necessarily orthogonal. To avoid correlation between movement along different axes, and to avoid the particle remaining at rest for an entire time step, we choose these operators to obey the conditions \cite{MlodinowBrun21,BrunMlodinow25}:
\begin{eqnarray}
\label{eq:equalnorm}
P_i^k P_j^+ P_i^k &=& P_i^k P_j^- P_i^k = c P_i^k , \nonumber\\
P_i^k P_j^0 P_i^k &=& c' P_i^k , \\
P_i^0 P_j^k P_i^0 &=& c' P_i^0 , \nonumber\\
P_X^0 P_Y^0 = P_Y^0 P_Z^0 &=& P_Z^0 P_X^0 = 0 , \nonumber
\end{eqnarray}
where $k=\pm$, $i,j=X,Y,Z$, $i\ne j$, and $c$ and $c'$ are some positive real constants.

Since we would like our solutions to correspond to photons in the long-wavelength limit, we would like the internal states to include two positive-energy solutions, corresponding ultimately to the two helicity states; by time-reversal symmetry, we will also have two negative energy solutions \cite{BrunMlodinow25}; and because the 0 modes enable us to get transversal solutions, we also need two 0 states. So the internal space of the particle in the QW is six-dimensional. We can construct a set of operators that satisfy the conditions in Eq.~(\ref{eq:equalnorm}) from the Pauli matrices for spin-$1/2$ particles,
\begin{equation}
\sigma_X = \left(\begin{array}{cc} 0 & 1 \\ 1 & 0 \end{array}\right) , \ \ \ 
\sigma_Y = \left(\begin{array}{cc} 0 & -i \\ i & 0 \end{array}\right) , \ \ \ 
\sigma_Z = \left(\begin{array}{cc} 1 & 0 \\ 0 & -1 \end{array}\right) ,
\end{equation}
and the $J$ matrices for spin-$1$ particles,
\begin{equation}
J_X = \left(\begin{array}{ccc} 0 & 0 & 0 \\ 0 & 0 & -i \\ 0 & i & 0 \end{array}\right) ,\ \ 
J_Y = \left(\begin{array}{ccc} 0 & 0 & i \\ 0 & 0 & 0 \\ -i & 0 & 0 \end{array}\right) ,\ \ 
J_Z = \left(\begin{array}{ccc} 0 & -i & 0 \\ i & 0 & 0 \\ 0 & 0 & 0 \end{array}\right) .
\label{eq:jMatrices}
\end{equation}
The $\sigma$ matrices have eigenvalues $\pm1$ and the $J$ matrices have eigenvalues $1,0,-1$. We can define a set of $6\times6$ matrices, in a manner analogous to the way the Dirac matrics are defined in terms of Pauli matrices in the fermion theory:
\begin{equation}
\gamma_j = \sigma_Z \otimes J_j ,\ \ \ j = X,Y,Z .
\label{eq:gammamatrices}
\end{equation}
The eigenspaces (with eigenvalues $1,0,-1$) of these matrices define the projectors $P_j^k$:
\begin{eqnarray}
P_j^+ &=& \frac{1}{2} (\gamma_j^2 + \gamma_j) , \nonumber\\
P_j^0 &=& I - \gamma_j^2 , \\
P_j^+ &=& \frac{1}{2} (\gamma_j^2 + \gamma_j) . \nonumber
\end{eqnarray}
It is easy to check that these form a set of orthogonal projectors for each $j=X,Y,Z$, and that they satisfy the conditions in Eq.~(\ref{eq:equalnorm}) with constants $c=1/4$ and $c'=1/2$.

The evolution operator in Eq.~(\ref{eq:bosonQW}) is manifestly unitary, so this gives a well-defined quantum walk. The state evolves in discrete steps of size $\Delta t$ by being multiplied by $U_{QW}$:
\[
\ket{\psi(t+\Delta t)} = U_{QW}\ket{\psi(t)} .
\]
In the next subsection we will show that the eigenstates of this evolution are discrete, plane-wave-like solutions.

\subsection{Solutions to the QW}
\label{sec:QWsolns}

When the linear size $N$ of the lattice is finite we can define a set of momentum eigenstates by taking the discrete Fourier transform of the position basis vectors:
\begin{equation}
\ket{\mathbf{x}} = \frac{1}{N^{3/2}} \sum_{\mathbf{k}} e^{-i\mathbf{x}\cdot\mathbf{k}} \ket{\mathbf{k}} ,
\ \ \ 
\ket{\mathbf{k}} = \frac{1}{N^{3/2}} \sum_{\mathbf{x}} e^{i\mathbf{x}\cdot\mathbf{k}} \ket{\mathbf{x}} .
\label{eq:momentumstates1}
\end{equation}
Here, the components $k_{x,y,z}$ of $\mathbf{k}$ are confined within the range $-\pi/\Delta x < k_{x,y,z} \le \pi/\Delta x$, and they are integer multiples of $\Delta k = 2\pi/N\Delta x$. (Here we have assumed that $N$ is even, so the components take values of the form $\ell\Delta k$ where $-N/2 < \ell \le N/2$.)

The key to solving for the eigenstates of $U_{QW}$ is to note that these momentum operators are eigenstates of the shift operators:
\begin{equation}
S_j \ket{\mathbf{k}} = e^{-i k_j \Delta x} \ket{\mathbf{k}} ,\ \ \ j=X,Y,Z .
\end{equation}
So if we have a state $\ket\Psi = \ket{\mathbf{k}}\otimes\ket{\phi}$, where $\ket{\phi}$ is a state in the particle's internal space, then applying $U_{QW}$ gives us
\begin{eqnarray}
U_{QW}\ket\Psi &=& \ket{\mathbf{k}} \otimes \left( e^{-i k_x\Delta x} P_X^+ + P_X^0 + e^{i k_x\Delta x} P_X^- \right) \nonumber\\
&& \times \left( e^{-i k_y\Delta x} P_Y^+ + P_Y^0 + e^{i k_y\Delta x} P_Y^- \right) \nonumber\\
&& \times \left( e^{-i k_z\Delta x} P_Z^+ + P_Z^0 + e^{i k_x\Delta x} P_Z^- \right) \ket\phi \nonumber\\
&=& \ket{\mathbf{k}} \otimes e^{-i k_x\Delta x\gamma_X} e^{-i k_y\Delta x\gamma_Y} 
e^{-i k_z\Delta x\gamma_Z} \ket\phi .
\end{eqnarray} 
The evolution unitary leaves the spatial part of the wavefunction $\ket{\mathbf{k}}$ unchanged, and acts on the internal space like the $6\times6$ unitary matrix
\begin{equation}
U_{\mathbf{k}} = e^{-i k_x\Delta x\gamma_X} e^{-i k_y\Delta x\gamma_Y} 
e^{-i k_z\Delta x\gamma_Z} .
\end{equation}
Using the definitions of the $\gamma$ matrices from Eq.~(\ref{eq:gammamatrices}) we can find this matrix explicitly:
\begin{equation}
U_{\mathbf{k}} = \left( \begin{array}{cccccc}
c_y c_z & -c_y s_z & s_y & 0 & 0 & 0 \\
c_z s_x s_y + c_x s_z & c_x c_z - s_x s_y s_z & - c_y s_x & 0 & 0 & 0 \\
-c_x c_z s_y + s_x s_z & c_z s_x + c_x s_y s_z & c_x c_y & 0 & 0 & 0 \\
0 & 0 & 0 & c_y c_z & c_y s_z & - s_y \\
0 & 0 & 0 & c_z s_x s_y - c_x s_z & c_x c_z + s_x s_y s_z & c_y s_x \\
0 & 0 & 0 & c_x c_z s_y + s_x s_z & -c_z s_x + c_x s_y s_z & c_x c_y
\end{array}\right) ,
\label{eq:ukmatrix}
\end{equation}
where $c_{x,y,z} = \cos(k_{x,y,z}\Delta x)$ and $s_{x,y,z} = \sin(k_{x,y,z}\Delta x)$. The eigenvalues of this matrix $U_{\mathbf{k}}$ are
\[
e^{-i \phi(\mathbf{k}) } , 1 , e^{+i \phi(\mathbf{k})} ,
\]
where $\phi(\mathbf{k})>0$; we refer to these eigenstates as positive-, zero-, and negative-energy solutions, respectively. They are each doubly degenerate; we interpret the two degenerate eigenstates as different helicity states. We denote these eigenstates as follows:
\begin{equation}
\ket{+\uparrow}_{\mathbf{k}} ,\ 
\ket{+\downarrow}_{\mathbf{k}} ,\ 
\ket{0\uparrow}_{\mathbf{k}} ,\ 
\ket{0\downarrow}_{\mathbf{k}} ,\ 
\ket{-\uparrow}_{\mathbf{k}} ,\ 
\ket{-\downarrow}_{\mathbf{k}} ,
\end{equation}
where $+,-,0$ refer to positive-, negative-, and zero-energy states, respectively, and $\uparrow$ and $\downarrow$ to the two helicity states. The subscript $\mathbf{k}$ is necessary because these eigenstates are different for different vectors $\mathbf{k}$. So a positive-energy eigenstate of $U_{QW}$, for example, would be a linear combination of $\ket{\mathbf{k}} \otimes \ket{+\uparrow}_{\mathbf{k}}$ and $\ket{\mathbf{k}} \otimes \ket{+\downarrow}_{\mathbf{k}}$. We refer to states written in terms of the eigenbasis of $U_{QW}$ as being in the ``energy'' representation. (Since time is discrete, we cannot consider this as a true energy, but in the long-wavelength limit we essentially recover the usual energy interpretation, with $\phi(\mathbf{k})$ playing the role of the Hamiltonian.)

If we take the limit $N\rightarrow\infty$, keeping the lattice spacing $\Delta x$ fixed, the momentum becomes continuous, but its components $k_{x,y,z}$ are still confined within the range $-\pi/\Delta x < k_{x,y,z} < \pi/\Delta x$. The momentum eigenstates $\ket{\mathbf{k}}$ are no longer normalizable, but instead have inner products
\[
\bracket{\mathbf{k}'}{\mathbf{k}} = \delta(\mathbf{k}-\mathbf{k}') .
\]
In this limit the position and moomentum eigenstates are related by
\begin{equation}
\ket{\mathbf{x}} = \left(\frac{\Delta x}{2\pi}\right)^{3/2} \int d^3\mathbf{k}\, e^{-i\mathbf{k}\cdot\mathbf{x}} \ket{\mathbf{k}} ,
\ \ \ 
\ket{\mathbf{k}} = \left(\frac{\Delta x}{2\pi}\right)^{3/2} \sum_{\mathbf{x}} e^{i\mathbf{k}\cdot\mathbf{x}} \ket{\mathbf{x}} .
\label{eq:momentumstates2}
\end{equation}

\subsection{Wave packets and group velocity}
\label{sec:wavepackets}

Now that we have found plane-wave solutions to the QW, let us look at discrete wave packets in the 3-dimensional lattice. In the position representation we have basis states of the form
\begin{equation}
\ket{\mathbf{x},\sigma} \equiv \ket{\mathbf{x}} \otimes \ket\sigma,
\end{equation}
where $\sigma=0,1,2,3,4,5$ and the possible positions are
\begin{equation}
\mathbf{x} = \left(\begin{array}{c} \ell \\ m \\ n \end{array}\right) \Delta x ,\ \ \ \ell,m,n\in\mathbb{Z} .
\end{equation}
If the lattice is finite with linear dimension $N$ then we can take the coordinates to lie within a range $-N/2 < \ell,m,n \le N/2$, but we can also take the limit $N\rightarrow\infty$ when it is convenient (even if the lattice is actually finite, but very large). Each step of the evolution takes time $\Delta t$, and we define the ``speed of light'' $c=\Delta x/\Delta t$.

In the ``energy'' representation, our basis states take the form
\begin{equation}
\ket{\mathbf{k},\eta} \equiv \ket{\mathbf{k}} \otimes \ket{\eta}_{\mathbf{k}} ,
\end{equation}
where $\eta=\pm\uparrow,\pm\downarrow,0\uparrow$ or $0\downarrow$, and the states $\ket{\eta}_{\mathbf{k}}$ depend on the momentum $\mathbf{k}$. These states $\ket{\mathbf{k},\eta}$ are eigenstates of the evolution unitary $U$:
\begin{equation}
U \ket{\mathbf{k},\eta} = \begin{cases} e^{-i \phi(\mathbf{k})} \ket{\mathbf{k},\eta} , & \eta=+\uparrow \text{ or } +\downarrow,\\
e^{+i \phi(\mathbf{k})} \ket{\mathbf{k},\eta} , &\eta=-\uparrow \text{ or } -\downarrow,\\
\ket{\mathbf{k},\eta} , & \eta=0\uparrow \text{ or } 0\downarrow .\end{cases} 
\end{equation}
These three cases denote positive-, negative- and zero-energy states, respectively. The Hamiltonian has two zero-energy longitudinal eigenstates, and a positive- and negative-energy eigenstate for each circular polarization. This is analogous to the situation in the theory of electrons and positrons, but here the photon is its own antiparticle, so the negative-energy solutions do not indicate the existence of additional particles, and in the free theory they do not mix. We will restrict ourselves to using positive-energy solutions. For more on this topic see \cite{BrunMlodinow25}.

The components of the momentum vector $\mathbf{k}$ lie within the range
\[
-\pi/\Delta x < k_{x,y,z} < \pi/\Delta x .
\]
If the lattice is finite with linear size $N$, then the $\mathbf{k}$ vectors are discrete,
\begin{equation}
\mathbf{k} = \left(\begin{array}{c} \ell \\ m \\ n \end{array}\right) \Delta k ,\ \ \ \Delta k = 2\pi/N\Delta x ,
\end{equation}
with $-N/2 < \ell,m,n \le N/2$. In the limit $N\rightarrow\infty$ the momenta become continuous, but remain bounded within the finite range $-\pi/\Delta x < k_{x,y,z} \le \pi/\Delta x$.

Now we will look at the evolution of a localized wave packet. Suppose we have a superposition of positive-energy eigenstates in a narrow region $R$ centered on a central momentum $\mathbf{k}_0$, of the form
\begin{equation}
\ket\psi = \sum_{\mathbf{k}\in R} \alpha_{\mathbf{k}} e^{-i \mathbf{k}\cdot\mathbf{x}_0} \ket{\mathbf{k},+\uparrow} ,
\label{eq:packet}
\end{equation}
where $\mathbf{x}_0$ is the center of the wave packet in the position representation, and we have arbitrarily chosen helicity to be $\uparrow$.

For now, we will assume that the momenta are confined in a finite region $R$, and not worry about the form of the coefficients $\alpha_{\mathbf{k}}$; but this argument goes through even if $R$ includes all momenta, so long as $|\alpha_{\mathbf{k}}|$ falls off rapidly away from $\mathbf{k}_0$. Applying the evolution unitary $U$ to the state in Eq.~(\ref{eq:packet}), we get
\begin{equation}
U\ket\psi = \sum_{\mathbf{k}\in R} \alpha_{\mathbf{k}} e^{-i \mathbf{k}\cdot\mathbf{x}_0} 
e^{-i\phi(\mathbf{k})} \ket{\mathbf{k},+\uparrow} .
\end{equation}

We can expand the momentum $\mathbf{k}$ around $\mathbf{k}_0$:
\[
\mathbf{k} \equiv \mathbf{k}_0 + \mathbf{d} .
\]
Let $R$ be narrowly centered on $\mathbf{k}_0$, so $|\mathbf{d}| \ll |\mathbf{k}_0|$. Then we can expand the phase
\begin{eqnarray}
\phi(\mathbf{k}) &=& \phi(\mathbf{k}_0 + \mathbf{d})
= \phi(\mathbf{k}_0) + \frac{d\phi}{d\mathbf{k}}(\mathbf{k}_0) \cdot \mathbf{d} 
+ O(|\mathbf{d}|^2) \nonumber\\
&=& \phi(\mathbf{k}_0) + \frac{d\phi}{d\mathbf{k}}(\mathbf{k}_0) \cdot (\mathbf{k} - \mathbf{k}_0) 
+ O(|\mathbf{d}|^2) \nonumber\\
&\equiv& \left[ \phi(\mathbf{k}_0) - \frac{d\phi}{d\mathbf{k}}(\mathbf{k}_0) \cdot \mathbf{k}_0 \right]
+ \mathbf{v}_g \cdot \mathbf{k} \Delta t + O(|\mathbf{d}|^2) .
\end{eqnarray}
The first quantity, in the square brackets, depends only on $\mathbf{k}_0$, so it will be the same for every term in the superposition. In the second term we have introduced the {\it group velocity} of the wave packet:
\begin{equation}
\mathbf{v}_g \equiv \frac{1}{\Delta t} \frac{d\phi}{d\mathbf{k}}(\mathbf{k}_0) .
\label{eq:groupvelocity}
\end{equation}
We can approximate the evolved wave packet state as
\begin{equation}
U\ket\psi \approx e^{-i \left[ \phi(\mathbf{k}_0) - \mathbf{v}_g \cdot \mathbf{k}_0 \Delta t \right]}
\sum_{\mathbf{k}\in R} \alpha_{\mathbf{k}} e^{-i \mathbf{k}\cdot(\mathbf{x}_0 + \mathbf{v}_g\Delta t)}  
\ket{\mathbf{k},+\uparrow} .
\end{equation}
The phase factor in front of the sum is just a global phase of $\phi(\mathbf{k}_0) - \mathbf{v}_g \cdot \mathbf{k}_0 \Delta t$. The phase inside the sum shifts the center of the wave packet from $\mathbf{x}_0$ to $\mathbf{x}_0 + \mathbf{v}_g\Delta t$. So, at least approximately, the wave packet is propagating with velocity $\mathbf{v}_g$. The terms of $O(|\mathbf{d}|^2)$ that we are neglecting in the phase could cause the wave packet to change its shape (e.g, the wave packet could spread or change its envelope in other ways), but as long as $|\mathbf{d}| \ll |\mathbf{k}_0|$ and the second derivative of $\phi(\mathbf{k})$ is small this will happen slowly. Over the course of a single step, the wave packet is almost unchanged.

If we look at the long-wavelength limit we recover the usual behavior of photons. In that limit, we find that
\begin{equation}
\phi(\mathbf{k}) \approx \Delta x \sqrt{|\mathbf{k}|^2} \equiv k \Delta x ,
\end{equation}
which implies
\begin{equation}
\frac{d\phi}{d\mathbf{k}}(\mathbf{k}_0) \approx \Delta x \frac{\mathbf{k}}{k} \Rightarrow
\mathbf{v}_g = \frac{1}{\Delta t} \frac{d\phi}{d\mathbf{k}}(\mathbf{k}_0) \approx c \frac{\mathbf{k}}{k} .
\end{equation}
In the long-wavelength limit, the group velocity is the speed of light $c$ in the direction of $\mathbf{k}$. As we will see, the next corrections will not be negligible at shorter wavelengths (i.e., higher energies).

\subsubsection{The sinc wave packet}

For a first example, we consider a simple square pulse in momentum centered around $\mathbf{k}_0 = (k_{0x},k_{0y},k_{0z})$:
\begin{equation}
R = \{\mathbf{k} = (k_x,k_y,k_z) | k_{0j} - \frac{\ell}{2}\Delta k \le k_j \le k_{0j} + \frac{\ell}{2}\Delta k, j=x,y,z\} .
\end{equation}
Here, $\ell$ is an even integer much smaller than $N$. Our initial wave packet is
\begin{equation}
\ket\psi = \frac{1}{(\ell+1)^{3/2}} \sum_{\mathbf{k}\in R} e^{-i\mathbf{k}\cdot\mathbf{x}_0}
\ket{\mathbf{k},+\uparrow} .
\label{eq:squarepulse}
\end{equation}
If we assume that the range around $\mathbf{k}_0$ is small, then we can approximate
\begin{equation}
\ket{\mathbf{k},+\uparrow} = \ket{\mathbf{k}}\otimes\ket{+\uparrow}_{\mathbf{k}}
\approx \ket{\mathbf{k}}\otimes\ket{+\uparrow}_{\mathbf{k}_0} .
\label{eq:kapprox}
\end{equation}
Applying the evolution operator $U$ to this state gives us (up to a phase)
\begin{equation}
U\ket\psi \approx \frac{1}{(\ell+1)^{3/2}} \left( \sum_{\mathbf{k}\in R} e^{-i\mathbf{k}\cdot(\mathbf{x}_0 + \mathbf{v}_g\Delta t)} \ket{\mathbf{k}} \right) \otimes\ket{+\uparrow}_{\mathbf{k}_0} .
\end{equation}

We can now do a discrete Fourier transform to the position representation. Substituting the expression for $\ket{\mathbf{k}}$ from Eq.~(\ref{eq:momentumstates1}) into Eq.~(\ref{eq:squarepulse}), we get
\begin{eqnarray}
\ket\psi &\approx& \frac{1}{(N(\ell+1))^{3/2}} \sum_{\mathbf{x}} \left(
\sum_{\mathbf{k}\in R} e^{-i\mathbf{k}\cdot\mathbf{x}_0 + i\mathbf{k}\cdot\mathbf{x}}  \right)
\ket{\mathbf{x}}\otimes\ket{+\uparrow}_{\mathbf{k}_0} \nonumber\\
&=& \frac{1}{(N(\ell+1))^{3/2}} \sum_{\mathbf{x}} \left(
\sum_{\mathbf{k}\in R} e^{-i\mathbf{k}\cdot(\mathbf{x}_0 - \mathbf{x})} \right) 
\ket{\mathbf{x}}\otimes\ket{+\uparrow}_{\mathbf{k}_0} \nonumber\\
&=& \frac{1}{(N(\ell+1))^{3/2}} \sum_{\mathbf{x}}
\left( \sum_{k_x\in R_x} e^{-i k_x(x_0 - x)} \right) \left( \sum_{k_y\in R_y} e^{-i k_y(y_0 - y)} \right) \nonumber\\
&& \qquad \qquad \times \left( \sum_{k_z\in R_z} e^{-i k_z(z_0 - z)} \right) 
\ket{\mathbf{x}}\otimes\ket{+\uparrow}_{\mathbf{k}_0} \nonumber\\
&=& \frac{1}{(N(\ell+1))^{3/2}} \sum_{\mathbf{x}}
\frac{\sin\left(\frac{\pi(\ell+1)(x-x_0)}{N\Delta x}\right)}{\sin\left(\frac{\pi(x-x_0)}{N\Delta x}\right)}
\frac{\sin\left(\frac{\pi(\ell+1)(y-y_0)}{N\Delta x}\right)}{\sin\left(\frac{\pi(y-y_0)}{N\Delta x}\right)}
\frac{\sin\left(\frac{\pi(\ell+1)(z-z_0)}{N\Delta x}\right)}{\sin\left(\frac{\pi(z-z_0)}{N\Delta x}\right)} \nonumber\\
&& \qquad \qquad \qquad \qquad \times e^{i\mathbf{k}_0\cdot(\mathbf{x}-\mathbf{x}_0)} \ket{\mathbf{x}} \otimes \ket{+\uparrow}_{\mathbf{k}_0} .
\label{eq:sincsoln}
\end{eqnarray}
In the equations above, the components of the vectors are
\[
\mathbf{k} = \left(\begin{array}{c} k_{x} \\ k_{y} \\ k_z \end{array}\right) ,\ \ \ 
\mathbf{x} = \left(\begin{array}{c} x \\ y \\ z \end{array}\right) ,\ \ \ 
\mathbf{x}_0 = \left(\begin{array}{c} x_0 \\ y_0 \\ z_0 \end{array}\right) ,
\]
and the three regions $R_{x,y,z}$ are
\begin{eqnarray*}
R_x &=& \{ k_x | k_{0x} - \frac{\ell}{2}\Delta k \le k_x \le k_{0x} + \frac{\ell}{2}\Delta k, \\
R_y &=& \{ k_y | k_{0y} - \frac{\ell}{2}\Delta k \le k_y \le k_{0y} + \frac{\ell}{2}\Delta k, \\
R_z &=& \{ k_z | k_{0z} - \frac{\ell}{2}\Delta k \le k_z \le k_{0z} + \frac{\ell}{2}\Delta k .
\end{eqnarray*}

As $N\rightarrow\infty$ the solution in Eq.~ (\ref{eq:sincsoln}) goes to a product of sinc functions,
\[
\mathrm{sinc}(\theta) \equiv \frac{\sin(\theta)}{\theta} .
\]
This is a very delocalized kind of wave packet in space---not surprising, since it is so sharply confined in momentum---but it does propagate with velocity $\mathbf{v}_g$, as desired.

\subsubsection{Approximate Gaussian wave packets}

The most commonly used wave packet definition in continuous space is the Gaussian wave packet. If we let $N\rightarrow\infty$ then $\mathbf{k}$ becomes continuous and periodic with period $2\pi/\Delta x$ in all three dimensions. We can then define an approximately Gaussian wave packet in $\mathbf{k}$:
\begin{equation}
\ket\psi = \frac{1}{(2\pi\sigma^2)^{3/4}} \int d^3\mathbf{k}\, e^{- {|\mathbf{k}-\mathbf{k}_0|}^2 /4\sigma^2
- i\mathbf{k}\cdot\mathbf{x}_0} \ket{\mathbf{k},+\uparrow} .
\end{equation}
It is only an approximate Gaussian because the tails do not really extend to $\pm\infty$, since $\mathbf{k}$ is periodic. This would also mean that the normalization above is not quite correct, but the difference will be very, very small if the width $\sigma$ is much smaller than the full range $2\pi/\Delta x$.

Since this wave packet has tails that extend throughout the entire range of $\mathbf{k}$, it is no longer true that $\mathbf{d} = \mathbf{k} - \mathbf{k}_0$ is small everywhere that the wave packet is nonvanishing. Nor can we approximate the internal state $\ket{+\uparrow}_{\mathbf{k}} \approx \ket{+\uparrow}_{\mathbf{k}_0}$ everywhere. However, if the Gaussian is narrow, then these deviations only matter on the tails, which have extremely low amplitude. So we can still approximate
\begin{equation}
U\ket\psi \approx \frac{1}{(2\pi\sigma^2)^{3/4}} \left( \int d^3\mathbf{k} e^{- {|\mathbf{k}-\mathbf{k}_0|}^2 /4\sigma^2 - i\mathbf{k}\cdot(\mathbf{x}_0 + \mathbf{v}_g\Delta t)} \ket{\mathbf{k}} \right) \otimes \ket{+\uparrow}_{\mathbf{k}_0} .
\label{eq:upsigaussian}
\end{equation}
Again, we can transform to the position representation, using the definitions in Eq.~(\ref{eq:momentumstates2}):
\begin{eqnarray}
\ket\psi &\approx& \frac{1}{(2\pi\sigma^2)^{3/4}} \int d^3\mathbf{k}\, 
e^{- {|\mathbf{k}-\mathbf{k}_0|}^2 /4\sigma^2 - i\mathbf{k}\cdot\mathbf{x}_0} 
\ket{\mathbf{k}} \otimes \ket{+\uparrow}_{\mathbf{k}_0} \nonumber\\
&=& \left(\frac{\Delta x}{\sigma}\right)^{3/2} \frac{1}{(2\pi)^{9/4}} \int d^3\mathbf{k}\, 
e^{- {|\mathbf{k}-\mathbf{k}_0|}^2 /4\sigma^2 - i\mathbf{k}\cdot\mathbf{x}_0} 
\sum_{\mathbf{x}} e^{i\mathbf{k}\cdot\mathbf{x}} \ket{\mathbf{x}} \otimes \ket{+\uparrow}_{\mathbf{k}_0} \nonumber \\
&=& \left(\frac{\Delta x}{\sigma}\right)^{3/2} \frac{1}{(2\pi)^{9/4}} \sum_{\mathbf{x}} 
\left( \int d^3\mathbf{k}\, e^{- {|\mathbf{k}-\mathbf{k}_0|}^2 /4\sigma^2
- i\mathbf{k}\cdot(\mathbf{x}_0-\mathbf{x})} \right) \ket{\mathbf{x}} \otimes \ket{+\uparrow}_{\mathbf{k}_0} \nonumber\\
&\approx& \left(\frac{\sigma^2}{8\pi}\right)^{3/4} (\Delta x)^{3/2} \sum_{\mathbf{x}} 
e^{- \sigma^2 {|\mathbf{x}-\mathbf{x}_0|}^2 + i\mathbf{k}_0\cdot(\mathbf{x} - \mathbf{x}_0)} 
\ket{\mathbf{x}} \otimes \ket{+\uparrow}_{\mathbf{k}_0} .
\end{eqnarray}
The first approximation is in using $\ket{+\uparrow}_{\mathbf{k}} \approx \ket{+\uparrow}_{\mathbf{k}_0}$, which is true where the amplitude is non-negligible, and the last approximation is in integrating from $-\infty$ to $\infty$, rather than $-\pi/\Delta x$ to $\pi/\Delta x$, which is a very good approximation when $1/\sigma \gg \Delta x$. The result, to a very good approximation, is a discrete Gaussian wave packet in the position description, centered at $\mathbf{x}_0$ and with average momentum $\mathbf{k}_0$. Evolving this wave packet one step as in Eq.~(\ref{eq:upsigaussian}) yields (up to a phase)
\begin{equation}
U\ket\psi \approx  \left(\frac{\sigma^2}{8\pi}\right)^{3/4} (\Delta x)^{3/2} \sum_{\mathbf{x}} 
e^{- \sigma^2 {|\mathbf{x}-\mathbf{x}_0|}^2 
+ i\mathbf{k}_0\cdot(\mathbf{x} - \mathbf{x}_0 - \mathbf{v}_g\Delta t)} 
\ket{\mathbf{x}} \otimes \ket{+\uparrow}_{\mathbf{k}_0} .
\end{equation}
So the wave packet propagates with velocity $\mathbf{v}_g$ while approximately maintaining its shape.

\subsection{LIV and anisotropy}
\label{sec:QCAliv}

In Sec.~\ref{sec:QWsolns} we derived the $6\times6$ matrix $U_{\mathbf{k}}$ in Eq.~(\ref{eq:ukmatrix}) that represents the effect of the evolution unitary $U_{QW}$ on the internal space of a single particle with exact momentum $\mathbf{k}$. The eigenvalues of this matrix give the phases associated with the different ``energy states'' of the particle. The twice-degenerate eigenvalues are:
\begin{eqnarray}
\lambda_+ &=& \frac{1}{2}\left( c_x c_y + c_x c_z + c_y c_z + s_x s_y s_z - 1 \right) \nonumber\\
&& - i \sqrt{1 - \frac{1}{4}\left( c_x c_y + c_x c_z + c_y c_z + s_x s_y s_z - 1 \right)^2} , \nonumber\\
\lambda_0 &=& 1 ,\\
\lambda_- &=& \frac{1}{2}\left( c_x c_y + c_x c_z + c_y c_z + s_x s_y s_z - 1 \right) \nonumber\\
&& + i \sqrt{1 - \frac{1}{4}\left( c_x c_y + c_x c_z + c_y c_z + s_x s_y s_z - 1 \right)^2} , \nonumber
\end{eqnarray}
recalling again the shorthand notation $c_{x,y,z} = \cos(k_{x,y,z}\Delta x)$ and $s_{x,y,z} = \sin(k_{x,y,z}\Delta x)$. We see, as stated in Sec.~\ref{sec:QWsolns}, that these eigenvalues have the form
\[
\lambda_+ = e^{-i\phi(\mathbf{k})} ,\ \ \ 
\lambda_0 = 1 = e^0 ,\ \ \ 
\lambda_- = e^{i\phi(\mathbf{k})} ,
\]
where the phase is
\begin{equation}
\phi(\mathbf{k}) = \cos^{-1}\left( (c_x c_y + c_x c_z + c_y c_z + s_x s_y s_z - 1)/2 \right) .
\label{eq:masslessbosonphase}
\end{equation}
This phase plays the role of a Hamiltonian; the corresponding energy would be $E(\mathbf{k}) = \hbar \phi(\mathbf{k})/\Delta t$. In the long-wavelength limit, we can expand the expression in Eq.~(\ref{eq:masslessbosonphase}) to get
\begin{equation}
\phi(\mathbf{k}) = k\Delta x - \frac{k_x k_y k_z \Delta x^2}{k} + O((k\Delta x)^3) ,
\label{eq:energyexpansion}
\end{equation}
where $k = |\mathbf{k}| = \sqrt{k_x^2 + k_y^2 + k_z^2}$. To leading order this is the same as the usual phase for a propagating photon.

As shown in the previous subsection, a wave packet propagates approximately at the group velecity $\mathbf{v}_g = (1/\Delta t)d\phi/d\mathbf{k}$, defined in Eq.~(\ref{eq:groupvelocity}). We can actually calculate this from the expression in Eq.~(\ref{eq:masslessbosonphase}):
\begin{eqnarray}
v_x &=& \frac{\Delta x}{\Delta t} \frac{s_x(c_y+c_z) - c_x s_y s_z}{\sqrt{4 - \left( c_x c_y + c_x c_z + c_y c_z + s_x s_y s_z - 1 \right)^2}}, \nonumber\\
v_y &=& \frac{\Delta x}{\Delta t} \frac{s_y(c_x+c_z) - s_x c_y s_z}{\sqrt{4 - \left( c_x c_y + c_x c_z + c_y c_z + s_x s_y s_z - 1 \right)^2}}, \nonumber\\
v_z &=& \frac{\Delta x}{\Delta t} \frac{s_z(c_x+c_y) - s_x s_y c_z}{\sqrt{4 - \left( c_x c_y + c_x c_z + c_y c_z + s_x s_y s_z - 1 \right)^2}} .
\end{eqnarray}

Here, the speed of light corresponds to $c = \Delta x/\Delta t$. It is not hard to see that this group velocity will not be exactly equal to $c$ for all momenta $\mathbf{k}$; and moreover, that the dependence of velocity on momentum will be anisotropic---it will depend significantly on the direction $\mathbf{\hat{k}} = \mathbf{k}/k$. However, this effect is strong only for large momenta. For small momenta, corresponding to the long-wavelength limit (where the wavelength is much longer than $\Delta x$), we can expand the components of $\mathbf{v}_g$ in the components of momentum:
\begin{eqnarray}
v_x &=& c\left(\frac{k_x}{k} + \frac{k_y k_z (k^2 - k_x^2)\Delta x}{k^3} + O((k\Delta x)^2) \right) , \nonumber\\
v_y &=& c\left(\frac{k_y}{k} + \frac{k_x k_z (k^2 - k_y^2)\Delta x}{k^3} + O((k\Delta x)^2) \right) , \nonumber\\
v_z &=& c\left(\frac{k_z}{k} + \frac{k_x k_y (k^2 - k_z^2)\Delta x}{k^3} + O((k\Delta x)^2) \right) .
\label{eq:vcomponents}
\end{eqnarray}
To leading order, the group velocity of a wave packet is exactly the speed of light $c$, in all directions---in the long-wavelength limit, the propagating waves are insensitive to the discreteness of the lattice. However, we also see that there will be a small deviation which does depend on both the magnitude and the direction of the momentum. If we calculate the magnitude of the group velocity, we get
\begin{equation}
v_g = |\mathbf{v}_g| = \sqrt{v_x^2 + v_y^2 + v_z^2}
= c \left( 1 - \frac{k_x k_y k_z \Delta x}{k^2} + O((k\Delta x)^2) \right) .
\label{eq:cdeviation1}
\end{equation}
Again, to leading order all wave packets propagate with velocity $c$ in all directions; but there is a correction of order $k\Delta x$ which can be either positive (superluminal) or negative (subluminal), and depends on the direction of the momentum. Intriguingly, from Eq.~(\ref{eq:energyexpansion}) we see that the speed of light is closely related to the phase $\phi(\mathbf{k})$ in the long wavelength limit:
\[
v_g = c \frac{\phi(\mathbf{k})}{k\Delta x} + O((k\Delta x)^2) .
\]

One minor clarification may be helpful. The deviation of the speed of light given in Eq.~(\ref{eq:cdeviation1}) is expressed in terms of the components of $\mathbf{k}$, which are not directly observable. We can see from Eq.~(\ref{eq:vcomponents}) that the vectors $\mathbf{k}$ and $\mathbf{v}_g$ are not exactly parallel; there is a small deflection of $\mathbf{v}_g/v_g$, of order $k\Delta x$, from the direction $\mathbf{\hat{k}}$. However, since $\mathbf{v}_g$ and $\mathbf{k}$ are equal to leading order, we can substitute $\mathbf{v}_g$ for $\mathbf{k}$ in Eq.~(\ref{eq:cdeviation2}) up to first order in $k\Delta x$:
\begin{equation}
v_g = c \left( 1 - (k\Delta x) \frac{v_x v_y v_z}{v_g^3} + O((k\Delta x)^2) \right) .
\label{eq:cdeviation2}
\end{equation}

We can represent the direction of $\mathbf{v}_g$ in spherical coordinates, in terms of two angular variables $\theta$ and $\phi$, as shown in Eq.~(\ref{eq:vspherical}). In terms of the angular variables the deviation of the speed of light becomes
\begin{equation}
v_g = c \left( 1 - (k\Delta x) \cos(\theta)\sin^2(\theta)\cos(\phi)\sin(\phi) + O((k\Delta x)^2) \right) .
\label{eq:cdeviation3}
\end{equation}
Comparing this to our expressions in Eq.~(\ref{eq:groupvelocity1}) and Eq.~(\ref{eq:groupvelocity2}), we see that the magnitude of the deviation depends on the energy of the photon and its direction,
\begin{equation}
v_g = c - s(\theta,\phi) \Delta v = c\left(1 - \frac{E}{E_{\mathrm{QG},1}} s(\theta,\phi) \right) ,
\end{equation}
where $\Delta v = ck\Delta x$, $E=\hbar ck$, $E_{\mathrm{QG},1}= c\hbar/\Delta x$ and
\begin{equation}
s(\theta,\phi) = \cos(\theta)\sin^2(\theta)\cos(\phi)\sin(\phi) .
\label{eq:angledependence}
\end{equation}
If we average $s(\theta,\phi)$ over all directions we get zero. So the average deviation in the speed of light is zero for all energies. However, the root-mean-square is nonzero:
\begin{equation}
\sqrt{\overline{s^2(\theta,\phi)}} = \sqrt{\frac{4\pi}{105}} \approx 0.346 .
\label{eq:RMSfactor}
\end{equation}
This gives us a baseline to compare the results of the QCA model to experimental observations.

\subsection{Bounds on lattice spacing}
\label{sec:bounds}

In Sec.~\ref{sec:observation} we detailed some observations that put experimental limits on the types of dispersive effects that are predicted by the QW model analyzed above, and hence also for the corresponding QCA model. There are two main effects that could be observable: (1) {\it dispersion}, in which the speed of light becomes frequency-dependent; and (2) {\it anisotropy}, in which the speed of light is different in different directions relative to the axes of the lattice. Both of these effects arise from Eq.~(\ref{eq:cdeviation3}). We should note that certain types of effects that have also been studied experimentally are {\it not} present in the QW model; for example, there is no birefringence, since the two helicity states are exactly degenerate with each other.

To some extent, each of these two effects complicates the observation of the other. The form of Eq.~(\ref{eq:cdeviation3}) includes both superluminal and subluminal deviations in different directions, with the deviations averaging to zero over all directions, so mixing together observations in different directions would not yield a consistent signal of dispersion. On the other hand, the deviations are energy-dependent, so that the anisotropy might be impossible to see except at the highest energies. These two effects together make it difficult to conclusively derive limits from experimental measurements. However, we can derive some probable upper bounds on the lattice spacing of the QW model.

To detect dispersion, observations from a single, high-energy astronomical event are best, since the light at all frequencies would arrive from the same direction. We have no way of knowing how these directions would line up with a hypothetical underlying lattice, so it is possible for an observation to be ``unlucky.'' If the direction of an event is within one of the planes corresponding to $v_x=0$, $v_y=0$ or $v_z=0$, the dispersion would be zero, and for directions close to such a plane it would be highly suppressed. So the best option would be to look at the limits derived separately from observations of multiple events in different directions.

The studies of dispersion in GRBs in \cite{Bolmont20,Cao24,LHAASO23}, discussed in Sec.~\ref{sec:LIVexpts}, are a good example. These studies put a limit on $E_{\mathrm{QG},1}$ in Eq.~(\ref{eq:dispersion}) of $E_{\mathrm{QG},1} \gtrsim 10^{20}$ GeV. Including the RMS factor in Eq.~(\ref{eq:RMSfactor}), we have the expression
\[
\frac{E_{\mathrm{QG},1}}{\sqrt{\overline{s^2(\theta,\phi)}}} = \frac{c\hbar}{\Delta x \sqrt{\overline{s^2(\theta,\phi)}}}
\]
from the model (on average), making this limit
\begin{equation}
\frac{c\hbar}{\Delta x \sqrt{\overline{s^2(\theta,\phi)}}} \gtrsim 10^{20}\,\mathrm{GeV} \ \Rightarrow\ 
\Delta x \lesssim \frac{c\hbar}{0.346\times 10^{20}\,\mathrm{GeV}} .
\end{equation}
We have $c\hbar\approx 2\times 10^{-7}$ eV-m, so to rough order this gives a limit of
\begin{equation}
\Delta x \lesssim 5.8\times 10^{-36}\, \mathrm{m} .
\end{equation}
For comparison, the Planck length is $L_P \approx 1.6\times 10^{-35}$ m. Intriguingly, this limit is within a little more than a factor of 2 of the Planck length.

As discussed in Sec.~\ref{sec:anisotropyexpts}, there are also experimental observations looking for anisotropy of the speed of light \cite{Mueller03,Hermann05,Antonini05,Hermann06,Herrmann09,Eisele09,Nagel15}. The experiments in \cite{Herrmann09,Eisele09,Nagel15} give a limit on an anisotropy of the speed of light of $\Delta c/c \lesssim 1\times 10^{-17}$--$10^{-18}$. For the QW model, the deviation in Eq.~(\ref{eq:cdeviation3}) differs most between the maximally positive and maximally negative directions. Given the angular dependence in Eq.~(\ref{eq:angledependence}) this maximum difference is
\begin{equation}
\Delta v_{\mathrm{max}} - \Delta v_{\mathrm{min}} \approx 0.385c\times \frac{E}{E_{\mathrm{QG},1}} .
\end{equation}
Taking the more stringent bound of $10^{-18}$, we get
\begin{equation}
\frac{\Delta c}{c} \approx 0.385\times \frac{E}{c\hbar/\Delta x} \lesssim 10^{-18} \ \Rightarrow\ 
\Delta x \lesssim 0.385\times 10^{-18}\times \frac{c\hbar}{E} .
\end{equation}
The energy of a photon is $E = \hbar\omega = \hbar(2\pi c/\lambda) = ch/\lambda$, where $\omega$ is the angular frequency and $\lambda$ is the wavelength. So the rough bound above becomes
\begin{equation}
\Delta x \lesssim 0.385\times 10^{-18}\times \lambda/2\pi .
\end{equation}
For the three experiments in question, two are done at near-infrared frequencies with $\lambda \approx 10^{-6}$ m, and one in the microwave range with $\lambda \approx 2\times 10^{-2}$ m. So the most stringent bound from these experiments is roughly
\begin{equation}
\Delta x \lesssim 6.5\times 10^{-26}\, \mathrm{m} ,
\end{equation}
which is about ten orders of magnitude weaker than the rough bound from GRBs. This difference is almost entirely due to the much lower energies of the photons in question. On the other hand, these experiments avoid the issue of being based on a relatively small number of events, whose orientation to the hypothetical background lattice, which strongly affects the strength of the dispersion, is unknown.

It is intriguing that the relatively simple QW model we consider, and the QCA models related to it, can recover relativistic and QFT equations in the long-wavelength limit. But it is purely a model of free particle propagation in a flat background spacetime. Even if it is possible to include interactions and develop QCA models that incorporate the entire Standard Model, how to incorporate gravity into such a theory is a huge open problem. However, some work in quantum gravity has explored the idea of a discrete structure for spacetime. (See, e.g., \cite{Penrose71,Markopoulou98,Baez00,Markopoulou04} and references therein.) It seems unlikely that such a theory would give a simple, regular lattice such as that which we considered; and irregular lattices might make it difficult to construct theories that support wave propagation, since irregularities can produce localized wave packet solutions, like in Anderson localization \cite{Anderson58,Keating07,Yin08}. If these difficulties could be overcome, however, we might expect results like those described in this paper.

\section{Conclusions}
\label{sec:conclusions}

Discrete-time quantum walks, elevated to quantum cellular automata (QCAs), reproduce the continuum limits of relativistic wave equations and even the free sector of quantum electrodynamics \cite{BrunMlodinow25}. This lattice-to-continuum map welds together two domains that historically evolved in parallel: the gate-based language of quantum information and the field-operator language of high-energy physics.

The synthesis naturally provokes the question of which description is more fundamental---does quantum field theory (QFT) emerge from an underlying digital substrate, or is the lattice merely an efficient numerical crutch for a truly continuous spacetime? If spacetime is fundamentally discrete, Lorentz symmetry would emerge as approximation, valid at long length scales. Because Lorentz invariance violation has not been observed, and because it is experimentally impossible to distinguish between continuous and discrete models if the underlying lattice spacing is small enough, current tests of Lorentz invariance at best impose an upper bound on the fundamental step size $\Delta x$. 

In the QCA realization of QED analyzed here, discreteness would reveal itself in two telltale ways:
\begin{enumerate}
\item Dispersion: the speed of light depends weakly on photon frequency;
\item Anisotropy: the velocity of light differs slightly along lattice axes versus oblique directions.
\end{enumerate}
Using arrival-time data from high-energy gamma-ray bursts (GRBs) \cite{Cao24,Bolmont20,LHAASO23} we inferred $\Delta x < 5.8\times10^{-36}$ m, within less than an order of magnitude of the Planck length $L_P \approx 1.6\times 10^{-35}$~m. Terrestrial Michelson–Morley-type resonator experiments \cite{Mueller03,Hermann05,Antonini05,Hermann06,Herrmann09,Eisele09,Nagel15}, which probe anisotropy with lower-energy photons, yield a weaker yet complementary bound of $\Delta x < 6.5\times 10^{-26}$~m, a gap of ten orders of magnitude that simply reflects the very different energy scales involved.

A confirmed violation of Lorentz symmetry would decisively favor a discrete picture; null results merely push the lattice spacing deeper into the ultraviolet. But if the observational bounds required the lattice spacing to be many orders of magnitude smaller than the Planck length, one could argue against QCAs based on a naturalness argument (i.e., dimensionless parameters in a theory should be of order 1). It is notable that this is not the case: the bounds allow the lattice spacing to be approximately what one might expect---the Planck length. 

This leaves two possibilities: either we happen to be at a time in which the limits of our technology by chance yield a bound at the Planck scale, or no better bound is possible because spacetime really is discrete at that scale. Prospects for sharper tests are excellent. Next-generation space-based gamma-ray observatories are poised to tighten dispersion and anisotropy limits \cite{Daniel15,Bozzo24}. In parallel, programmable quantum simulators may eventually be able to emulate QCA dynamics directly, enabling controlled tabletop probes of discrete-spacetime physics long before accelerators reach Planckian energies.

\acknowledgments

LM and TAB gratefully acknowledge the support of the Institute for Quantum Information and Matter at Caltech for hosting our collaboration for many years. This research was supported in part by NSF Grant PHYS-2310794

\end{document}